%% file: main.tex
\documentclass[conference]{IEEEtran}
\IEEEoverridecommandlockouts

\def\BibTeX{{\rm B\kern-.05em{\sc i\kern-.025em b}\kern-.08em
    T\kern-.1667em\lower.7ex\hbox{E}\kern-.125emX}}
\usepackage{cite}
\usepackage{amsmath,amssymb,amsfonts,amsthm}
\usepackage{algorithmic}
\usepackage{graphicx}
\usepackage{textcomp}
\usepackage{xcolor}
\usepackage{latexsym}

\usepackage{gensymb}
\usepackage{siunitx}
\usepackage{tikz}
\usepackage{etoolbox}
\usepackage{enumitem}
\usepackage{esvect}
\usepackage{mathtools}
\usetikzlibrary{chains,arrows,arrows.meta,calc,positioning}
\usepackage{multirow}
\usepackage{comment}
\usepackage[normalem]{ulem}
\usepackage{upquote}
\usepackage{tikz}
\usetikzlibrary{angles,quotes,positioning,calc,shapes.geometric, arrows.meta, shadows.blur}
\newcommand{\bnum}[1]{\textbf{\num{#1}}}

\input{packages}
\usepackage{dsfont}
\usepackage{booktabs}
\usepackage{tabularx}
\usepackage{siunitx}

\newcommand{\subsf}{\sf \scriptscriptstyle}
\newcommand{\bs}{\boldsymbol}
\newcommand{\mc}{\mathcal}
\newcommand{\wh}{\widehat}

\newcommand{\R}{\mathbb{R}}

\usepackage{cleveref}
\crefname{figure}{Fig.}{Figs.}
\Crefname{figure}{Fig.}{Figs.}
\crefname{table}{Tab.}{Tabs.}
\Crefname{table}{Tab.}{Tabs.}
\crefname{section}{Sec.}{Secs.}
\Crefname{section}{Section}{Sections}

\begin{document}

\title{LOCUS-DT: Localization via Observation-Conditioned Uncertainty Scoring with Digital Twins}

\author{
\IEEEauthorblockN{
Haozhe Lei$^{*1}$, 
Roberto Bomfin$^{2}$, 
Marwa Chafii$^{1,2}$,
and Sundeep Rangan$^{1}$
}
\IEEEauthorblockA{
$^1$NYU WIRELESS, Tandon School of Engineering, New York University, Brooklyn, NY 11201, USA\\
$^2$Engineering Division, New York University Abu Dhabi, 129188, UAE\\
hl4155@nyu.edu; roberto.bomfin@nyu.edu; marwa.chafii@nyu.edu; srangan@nyu.edu
}

\thanks{This work is supported in part by NSF grants 1952180, 2133662, 2236097, 2148293, and 1925079, the DARPA Prowess program, the NTIA, Tamkeen under the Research Institute NYUAD grant CG017, and industrial affiliates of NYU Wireless.}
\thanks{$^{*}$Corresponding author: Haozhe Lei (email: hl4155@nyu.edu).}
}

\maketitle


\begin{abstract}
Accurate indoor localization is essential for emerging applications in robotic navigation and search and rescue. While classical methods typically focus on single-point estimates, complex indoor environments with heavy blockage and multipath propagation often lead to multimodal likelihood surfaces where a single estimate is insufficient. This paper proposes LOCUS-DT (Localization via Observation-Conditioned Uncertainty Scoring with Digital Twins), a framework that treats snapshot localization as posterior inference over the transmitter location. By leveraging a ray-tracing-based digital twin (DT) of the known environment, LOCUS-DT generates synthetic multipath profiles for candidate locations and compares them against the measured channel profile. Central to our approach is a novel learned scoring function designed to compare a fixed number of dominant specular paths, providing robustness against errors in both the DT environment model and the physical channel estimation. Importantly, LOCUS-DT is trained over an ensemble of environments to ensure generalization to unseen layouts. We evaluate the system using a Sionna-based ray-tracing backend, demonstrating that LOCUS-DT captures the sharp, multimodal posterior structures inherent in indoor settings more accurately than standard Gaussian or Gaussian-mixture benchmarks.
\end{abstract}


\input{intro}

\input{problem}

\input{experiments}

\input{conclusion}

\bibliographystyle{IEEEtran}
\bibliography{ref}

\end{document}

%% file: intro.tex
\section{Introduction}

Wireless localization from a transmitted signal is a core capability in modern sensing and communication systems \cite{TR38.859,Rel18Enhance24}. It is also a key sensing function in emerging integrated sensing and communication systems. Classical localization methods typically return a single \emph{point estimate} for a transmitter's
location.
Indoor upper-mid-band measurements report rich angular dispersion \cite{Shakya25AngularSpread} and frequency-dependent delay spread \cite{Ying25IndoorFactory}, illustrating the multipath that makes single-snapshot localization ambiguous.  A wide range of downstream tasks, such as robotic navigation and search, require a more complete spatial map of the relative likelihood
of each possible transmitter location.  Probabilistically,
this problem can be seen as estimating a \emph{posterior distribution} on the transmitter's location given the measurements \cite{wang2025uncertainty,Lei25FullPosterior}. At the same time, many data-driven indoor localization methods remain closely tied to the environments and radio-frequency (RF) patterns observed during training, which can limit robustness when the layout-dependent propagation structure changes \cite{HeChan16,Wang15}.

\begin{figure*}[t]
    \centering
    \includegraphics[width=0.9\linewidth]{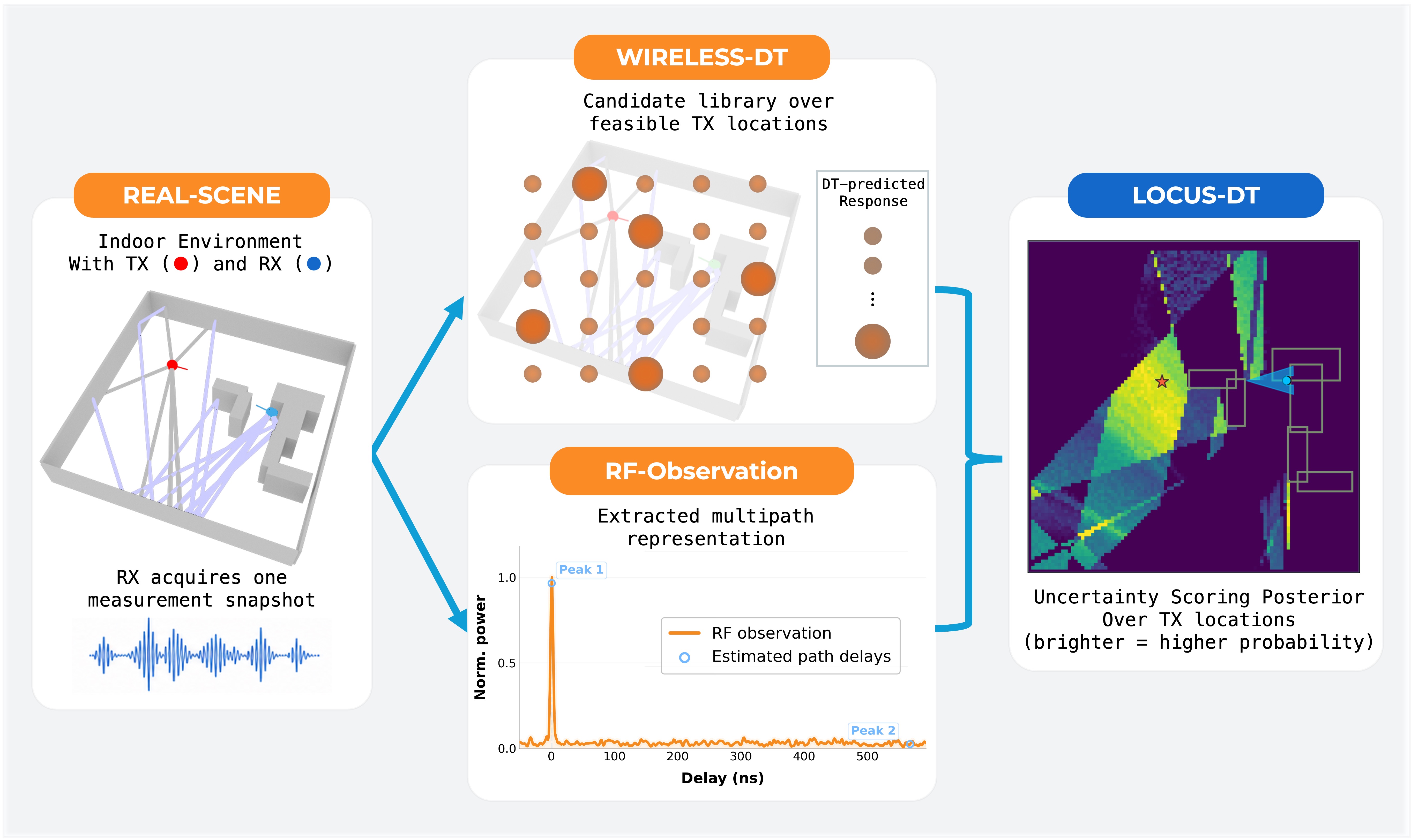}
    \caption{Concept of LOCUS-DT. From a real indoor scene, the receiver acquires one measurement snapshot. 
An offline wireless digital twin (WIRELESS-DT) module constructs a candidate library over feasible transmitter locations and generates the corresponding candidate-side multipath signatures.
The orange markers in the WIRELESS-DT panel denote candidate search locations; larger markers are used only to visually emphasize candidates with more salient DT-predicted multipath signatures. 
In parallel, an online RF-Observation module extracts a compact ranked multipath representation from the received signal, such as path-wise angle, delay, and strength features. 
LOCUS-DT then matches the observed multipath signature against the candidate-side DT signatures and outputs a posterior uncertainty heatmap over transmitter locations.}
    \label{fig:front}
\end{figure*}

In this work, we propose \emph{LOCUS-DT} (\emph{Localization via Observation-Conditioned Uncertainty Scoring with Digital Twins}), which casts indoor snapshot localization as posterior inference over the transmitter location given the measurements.
A conceptual figure is shown in \Cref{fig:front}.
The method assumes that the room layout and receiver (RX) pose are known.  Under this assumption,
the multipath profile from any candidate transmitter (TX) location can be estimated via a ray-tracing-based digital twin (DT).  At the same time, the measured multipath
profile can be estimated from the received signal.  By comparing the DT-generated and measured multipath profiles, each candidate location can be scored to obtain a posterior distribution over all candidate locations.

\subsection{Our Contributions}

\noindent
The main contributions of this work are as follows.

\vspace{2pt}\noindent
\textbf{DT-conditioned posterior formulation for indoor localization.}
We formulate single-snapshot indoor localization as posterior inference over the TX location given the layout and measurements.  The layout is incorporated implicitly through the ray-tracing DT to obtain multipath channel estimates for each candidate location.
These multipath channel estimates can be compared with the measured channel profile to score the likelihood of each candidate location.

\vspace{2pt}\noindent
\textbf{Learned scoring function.}
We propose a novel learned scoring function that
can compare the top $K$ paths of the candidate and measured multipath profiles.  Importantly, training can be performed over an ensemble of environments, so the method can be evaluated on environments not included in the training set.  Errors in either the DT or channel estimation can be incorporated during training for robustness.

\vspace{2pt}\noindent
\textbf{Validation with realistic ray tracing.}
We instantiate LOCUS-DT using a standard maximum-likelihood estimator of the specular components and a Sionna-based ray-tracing DT backend, and benchmark it against Gaussian and Gaussian-mixture posterior models. The experiments show that explicit candidate-wise DT matching better captures the sharp, multimodal, and layout-dependent posterior structure induced by indoor multipath.

\subsection{Prior Work}

Most RF localization methods are designed to produce point estimates, while the full spatial distribution over feasible transmitter locations is much less commonly modeled. Classical analyses such as the Cram\'er--Rao bound characterize estimation limits under assumed signal models \cite{godrich2008cramer}, and several recent works quantify uncertainty only through restricted summaries, such as output error scales, ranging uncertainty, or uncertainty in anchor and channel information \cite{zhang2024rloc,shamsfakhr2022indoor,muppirisetty2015spatial}. These approaches provide useful uncertainty characterizations, but they do not formulate indoor localization as posterior inference over a structured set of layout-conditioned transmitter hypotheses.

The work here builds on \cite{Lei25FullPosterior}, which estimates the posterior from a single dominant path.  Specifically, the current work augments \cite{Lei25FullPosterior} by introducing a DT to compute a full multipath profile from each candidate location and then comparing the DT multipath profile with an estimated or measured multipath profile.

At a broader level, our work is related to conditional density estimation, where a distribution is obtained by normalizing learned scores over a discretized output space \cite{silverman2018density,hall1999methods,lindsey1974comparison,gao2022lincde}. It is also related to recent efforts toward full-posterior wireless localization \cite{Lei25FullPosterior}. The present paper differs in its emphasis on layout-aware indoor inference: the normalized scores are not generic class logits, but candidate-wise compatibility scores between a measured multipath observation and a DT-predicted multipath signature.

Finally, significant prior work has studied wireless digital twins and their ray-tracing calibration and environment reconstruction \cite{khan2022digital,ying2026site,ying2026horama}, and has used ray tracing, digital twins, and physics-informed learning for sensing, localization, navigation, and downstream wireless decision-making \cite{ying2025isac,LeiICRA2024,Chen25TransformerRatePrediction,Lei2025DTWIN,Li25RLPhysics}. Our use of the DT is different in emphasis. Rather than using the DT only as a forward simulator or deterministic matching tool, we use it to generate a library of candidate-indexed multipath hypotheses and cast localization as posterior generation from candidate scores.

%% file: problem.tex
\section{LOCUS-DT Problem Formulation and Posterior Model}
\label{sec:prob_form}


\subsection{Localization Setting}
The TX is located at an unknown position $\bs{X}^t \in \R^d$ in a global coordinate system.  For simplicity, we take $d=2$,
so the problem is two-dimensional (2D) localization.
To estimate the TX location, the RX is given
\begin{equation}
\left(\mathcal L,\bs p\right),  \quad 
\bs p = (\bs x^r,\phi^r),
\label{eq:mirror_context}
\end{equation}
where $\mathcal L$ is a known room layout and
$\bs p$ is the receiver pose.  Since $d=2$, the pose is described by an RX position $\bs x^r \in \mathbb R^2$ and orientation
$\phi^r \in [-\pi,\pi)$.  At runtime, the RX attempts to listen for the signal from the TX and collects an observation $\bs y$.  Typically, the measurement $\bs{y}$ is an array of in-phase/quadrature (IQ) samples from the RX antennas; see the observation model in
\Cref{sec:online_obs}.  A \emph{point estimate}
would be a single vector $\wh{\bs{x}}^t$.  In contrast, the inference objective is to estimate the \emph{posterior distribution}
\begin{equation}
p\!\left(
\bs x^t
\mid
\mathcal L,\bs p,\bs y
\right),
\label{eq:mirror_target_posterior}
\end{equation}
representing the relative likelihood of all potential TX locations $\bs{x}^t$ given the observation, room layout, and receiver pose.

\subsection{DT Multipath Estimator}
The key structural assumption in LOCUS-DT is the availability of a DT that predicts the multipath signature generated by a hypothetical TX
location under a known room layout and receiver pose.
Formally, we describe each path as a tuple
\begin{equation}
    \bs \psi = (\theta,\tau,\alpha,\gamma) \in \mc{S},
\label{eq:mirror_generic_tuple}
\end{equation}
where $\theta$ is the angle of arrival (AoA), $\tau$ is the relative delay,
$\alpha \in \mathbb C$ is the complex path gain, and
$\gamma=|\alpha|^2/\sigma^2$ is its real-valued path SNR for receiver-noise
variance $\sigma^2$.  The set $\mc{S}$ contains such tuples.  The DT can then be modeled abstractly as a function:
\begin{equation}
\Psi(\mathcal L,\bs p,\bs x^t) = \left\{\bs\psi^{\subsf DT}_k, k=1,\ldots,K\right\}
\in \mc{S}^K,
\label{eq:mirror_dt_map}
\end{equation}
that maps each candidate location $\bs x^t$ to a predicted set of $K$ paths.  In our simulations below, $K$ is fixed to a small value, and the DT produces the top $K$ paths.
These quantities are retained for $K$ paths, sorted from strongest to weakest.
The abstraction is intentionally backend-agnostic.
Any DT method can be used to instantiate $\Psi(\mathcal L,\bs p,\bs x^t)$
as long as it produces the same ranked tuple representation in the common space
$\mc{S}$.

\subsection{Observation Interface}
\label{sec:online_obs}
The TX is assumed to periodically transmit a known waveform.  To search for the TX, the RX captures a snapshot of IQ samples, which we denote by
\begin{equation}
\bs y \in \mathbb C^{N \times N_{\subsf rx}},
\label{eq:mirror_raw_iq}
\end{equation}
where $N$ is the number of samples and $N_{\subsf rx}$ is the number of RX antennas.
We then model the multipath estimator as a function:
\begin{equation}
\wh{\Psi} : \mathbb C^{N \times N_{\subsf rx}} \rightarrow \mc{S}^K
\label{eq:mirror_estimator}
\end{equation}
which produces the estimated multipath profile
\begin{equation}
 \wh{\Psi} \left(\bs y\right)
=
\left\{
\bs \psi_{k}^{\subsf obs}, k=1,\ldots, K\right\}
\in \mc{S}^K.
\label{eq:mirror_observation}
\end{equation}
There is a large body of multipath estimators 
for path extraction from array measurements.
In this work, we use the iterative Levenberg--Marquardt (LM) algorithm \cite{Bomfin_multiband}.

\subsection{Score Function and Posterior} 
The core modeling problem is to construct a logit function that captures the
compatibility between the DT signature of a candidate location and the
multipath structure extracted from the real signal.
We formalize this process as finding a score function:
\begin{equation}
g_{\bs\omega}(\bs x^t,\mathcal L,\bs p,\bs y)
=
G_{\bs\omega}\!\left[
\{\bs \psi_{k}^{\subsf DT}(\bs{x}^t)\}, 
\{\bs \psi_{k}^{\subsf obs}\}
\right],
\label{eq:mirror_score_fn_pairs}
\end{equation}
where $\{\bs \psi_{k}^{\subsf DT}(\bs{x}^t)\}$
are the multipath components estimated by the DT in
\eqref{eq:mirror_dt_map} for the candidate TX location $\bs{x}^t$, and
$\{\bs \psi_{k}^{\subsf obs}\}$
is the observed multipath profile.
The vector $\bs\omega$ collects all trainable weights and biases of the neural network realizing the score function $G_{\bs\omega}$.

We next assume a prior distribution $p_0(\bs{x}^t)$,
which is typically uniform over a region of interest.
We then assume the posterior distribution is given by:
\begin{equation}
\hat p_{\bs\omega}\!\left(
\bs x^t
\mid
\mathcal L,\bs p,\bs y
\right)
=
\frac{
e^{g_{\bs\omega}(\bs x^t,\mathcal L,\bs p,\bs y)}\,p_{0}(\bs x^t)
}
{
Z_{\bs\omega}(\bs y;\mathcal L,\bs p)
},
\label{eq:mirror_posterior_estimator}
\end{equation}
where $Z_{\bs\omega}(\cdot)$ is the normalizing constant:
\begin{equation}
Z_{\bs\omega}(\bs y;\mathcal L,\bs p)
=
\int
e^{g_{\bs\omega}(\bs x^t,\mathcal L,\bs p,\bs y)}\,p_{0}(\bs x^t)
\;d\bs x^t.
\label{eq:mirror_partition}
\end{equation}

\begin{figure*}[!t]
  \centering
  \includegraphics[width=0.86\textwidth]{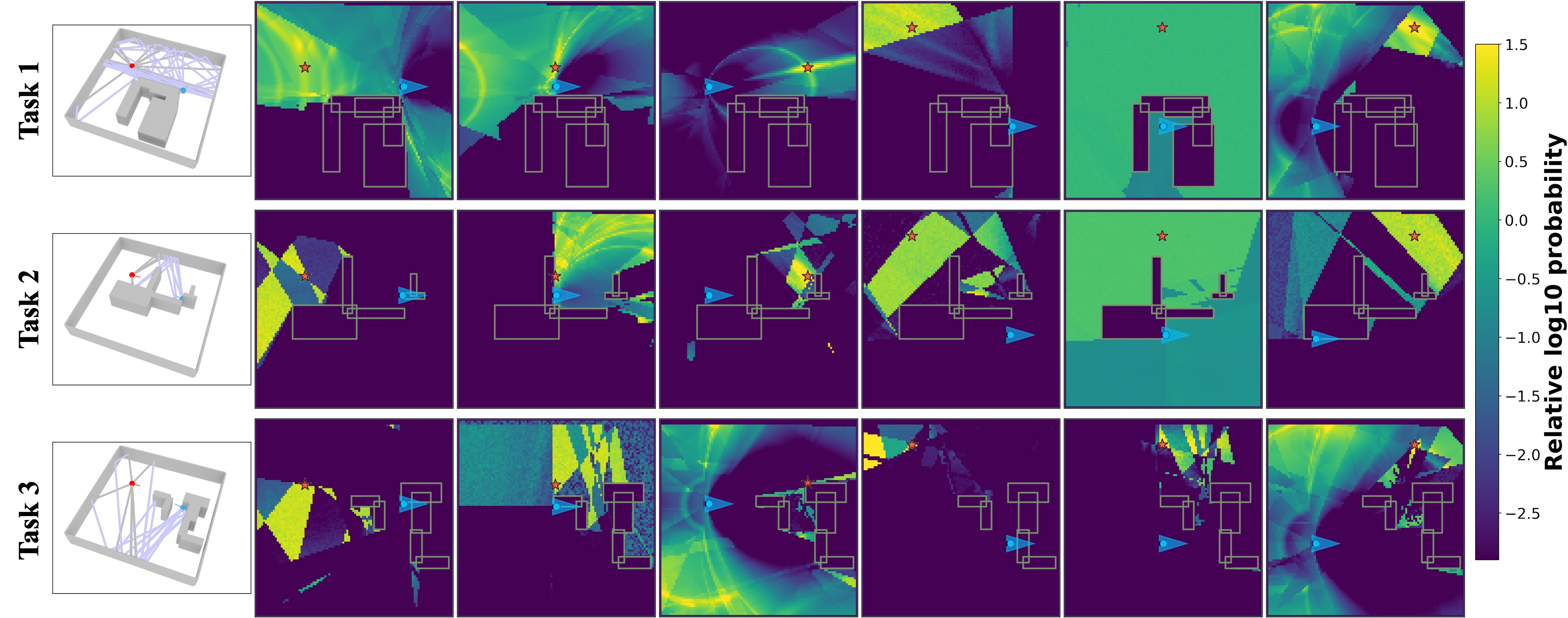}
  \caption{LOCUS-DT visualization for three tasks on distinct random indoor layouts.
  In each row (Task 1--3), the first panel shows the underlying Sionna room geometry,
  including the outer wall and interior obstacles, and the six panels to its
  right show posterior heatmaps for six corresponding TX/RX configurations on
  that same layout. The red five-point star marks the true TX location, the
  blue RX marker and arrow indicate the receiver position and orientation, and
  the outlined boxes indicate obstacles inside the room. Brighter colors
  correspond to larger relative log posterior values and therefore higher
  candidate likelihood.}
  \label{fig:dmc_fig1_direct}
\end{figure*}

\subsection{Sampled Cross-Entropy Loss}

The score function $G_{\bs\omega}[\cdot]$
can now be trained similarly to the earlier work
\cite{Lei25FullPosterior}, which does not use a DT.
As described in \cite{Lei25FullPosterior}, the key challenge in training a posterior of the form 
\eqref{eq:mirror_posterior_estimator} is that the partition function $Z_{\bs\omega}(\cdot)$ in \eqref{eq:mirror_partition} has no closed-form expression and involves numerical integration.
The partition function is instead replaced by a Monte Carlo estimate.
Specifically, we assume we are given a data set
\begin{equation}
\mathcal D
=
\left\{
\left(
\mathcal L_m,\bs p_m,\mathcal C_m,\bs y_m,\bs x_m^t
\right)
\right\}_{m=1}^{M},
\label{eq:mirror_dataset}
\end{equation}
where $M$ is the number of training samples.  
For each training sample, we are given a finite set
of \emph{candidate} TX locations:
\begin{equation}
\mathcal C_m = \{\bar{\bs x}_{mj}^t\}_{j=1}^{N_m} \subset \mathbb R^d.
\label{eq:mirror_dataset_candidates}
\end{equation}
The true TX location satisfies $\bs x_m^t \in \mathcal C_m$.  
Equivalently, there exists a unique candidate index
$c_m^\star \in \{1,\ldots,N_m\}$ such that
\begin{equation}
\bs x_m^t = \bar{\bs x}_{m c_m^\star}^{t}.
\label{eq:mirror_true_index}
\end{equation}
For each instance $m$, we evaluate the score function over the entire candidate
set $\mathcal C_m$ and define
\begin{equation}
g_{mj}
=
g_{\bs\omega}(\bar{\bs x}_{mj}^t,\mathcal L_m,\bs p_m,\bs y_m),
\qquad
j=1,\ldots,N_m.
\label{eq:mirror_dataset_logits}
\end{equation}
The ideal cross-entropy objective associated with
\eqref{eq:mirror_posterior_estimator} is
\begin{equation}
\mathcal L_{\mathrm{CE}}(\bs\omega)
=
-\frac{1}{M}
\sum_{m=1}^{M}
\log
\hat p_{\bs\omega}\!\left(
\bs x_m^t
\mid
\mathcal L_m,\bs p_m,\bs y_m
\right).
\label{eq:mirror_ce_true}
\end{equation}
In LOCUS-DT, we replace the integral partition function
\eqref{eq:mirror_partition} with the sampled candidate-set normalization
\begin{equation}
\hat Z_{\bs\omega}(\bs y_m;\mathcal L_m,\bs p_m)
=
\sum_{j=1}^{N_m}
e^{g_{mj}}.
\label{eq:mirror_partition_sampled_dataset}
\end{equation}
We then drop the additive constant induced by the uniform prior on $\mathcal C_m$.
Substituting this sampled normalization into \eqref{eq:mirror_ce_true} and using \eqref{eq:mirror_true_index}, \eqref{eq:mirror_dataset_logits}, and
\eqref{eq:mirror_partition_sampled_dataset},
we can approximate the loss function \eqref{eq:mirror_ce_true} as
\begin{equation}
\mathcal L(\bs\omega)
=
\frac{1}{M}
\sum_{m=1}^{M}
\left[
-g_{m c_m^\star}
+
\log\!\left(
\sum_{j=1}^{N_m} e^{g_{mj}}
\right)
\right].
\label{eq:mirror_ce_index}
\end{equation}
The loss in \eqref{eq:mirror_ce_index} is the
candidate-set analogue of the sampled cross-entropy criterion in energy-based
learning.

\subsection{Pairwise Neural Realization of the Score Function}

The score function $G_{\bs\omega}[\cdot]$ is realized
as a transformation followed by a simple multilayer perceptron.  In the simulations, we retain $K=6$
paths from each DT and observation profile.  Synchronization is unavailable and the system is narrowband, so we discard the delay.
Each AoA $\theta$ is encoded as $(\cos \theta,\sin\theta)$, while the SNR $\gamma$ is converted to dB and scaled to $[0,1]$ over a \SI{60}{dB} range.
The resulting $12$ paths yield $3\times 12 = 36$ features, which are fed into a two-layer neural network.  This architecture provides
an extremely compact, easily trainable scoring function with a limited number of parameters.

%% file: experiments.tex
\section{Experimental Instantiation of LOCUS-DT}
\label{sec:exp_inst}


\subsection{Synthetic Environment Generation}

To validate the methodology, we create random
synthetic environments with varying obstacles.
Importantly, the method is \emph{not} site-specific.
It can be tested on a completely random environment
different from those used for training.
Each random environment for training or testing is a $100~\mathrm{m}
\times 100~\mathrm{m}$ region.
Each scene contains the outer room boundary together with three randomly
generated interior obstacles drawn from rectangular, T-shaped, and L-shaped
primitives.
The NVIDIA Sionna ray-tracing backend \cite{sionna}
is then used to simulate the true channel and the $N_m = 500$ candidate channels.
\Cref{fig:dmc_fig1_direct} shows examples of these synthetic environments,
where the leftmost panel in each row gives a concrete room layout.

The carrier frequency is $12~\mathrm{GHz}$ and the signal bandwidth is
$200~\mathrm{MHz}$.
We use a noise figure of $7~\mathrm{dB}$, a synchronization window of
$4~\mu\mathrm{s}$, and a transmit power of $0~\mathrm{dBm}$.
The transmitter uses one vertically polarized isotropic element, while the
receiver uses a $1 \times 8$ vertically polarized half-wavelength array with a
3GPP TR~38.901 element pattern \cite{TR38.901}.
The ray tracer returns up to $16$ valid paths per channel; LOCUS-DT retains the $K=6$ strongest paths from each DT candidate profile.
For the online observation, the standard maximum-likelihood (SML) estimator uses a $128$-point matched filter to estimate $(\theta,\tau,\alpha)$, derives $\gamma$, and retains the $K=6$ strongest scorer pairs $(\theta,\gamma)$.

The training split therefore spans $8000$ distinct room layouts, each with
$500$ candidate transmitter locations and $10$ observation realizations, for a
total of $80{,}000$ snapshots.
The test split spans $2000$ additional layouts with the same
$500$-candidate structure, yielding $20{,}000$ snapshots.
For the final benchmark comparison, we report two harder evaluation sets:
Eval Random uses $2000$ held-out groups with $1000$ feasible random candidates
per group, while Eval Grid uses the corresponding held-out layouts but starts
from a regular $31 \times 31$ raw grid and keeps $696$ layout-feasible
candidate positions after pruning invalid points.

\begin{figure*}[t]
  \centering
  \includegraphics[width=0.86\textwidth]{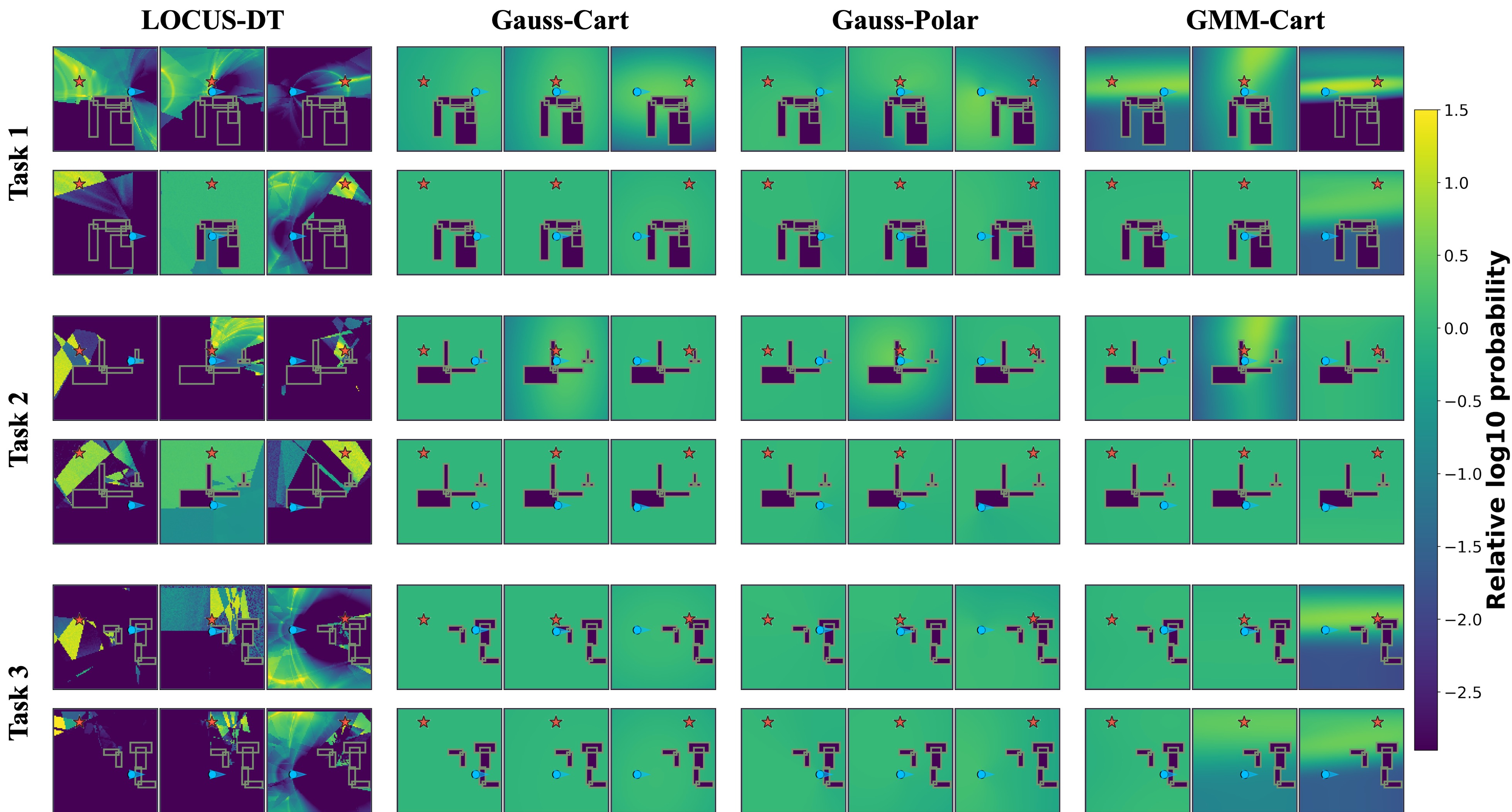}
  \caption{Qualitative comparison for the same three tasks and TX/RX
  configurations as in \Cref{fig:dmc_fig1_direct}.
  The visualization conventions are unchanged: the red five-point star marks
  the true TX, the blue marker and arrow indicate the RX pose, interior boxes
  indicate obstacles, and brighter colors correspond to larger relative log
  posterior values.}
  \label{fig:dmc_fig2_methods}
\end{figure*}

\begin{table*}[h]
  \centering
\caption{Adjusted loss $\mathcal{L}$ and derived metrics $\mathcal{G}$ and $\mathcal{R}$ on the indoor RF scenario. Eval Grid and Eval Random use 696 and 1000 candidates, respectively.}
\label{tab:dmc-eval-baselines-vs-mirror}
\resizebox{\textwidth}{!}{%
\begin{tabular}{llcccccccccccc}
  \toprule
  \multirow{2}{*}{Scenario} & \multirow{2}{*}{Evaluation Set} & \multicolumn{3}{c}{LOCUS-DT} & \multicolumn{3}{c}{Gauss-Cart} & \multicolumn{3}{c}{Gauss-Polar} & \multicolumn{3}{c}{GMM-Cart} \\
  \cmidrule(lr){3-5}\cmidrule(lr){6-8}\cmidrule(lr){9-11}\cmidrule(lr){12-14}
  &  & $\mathcal{L}$ & $\mathcal{G}$ & $\mathcal{R}$ & $\mathcal{L}$ & $\mathcal{G}$ & $\mathcal{R}$ & $\mathcal{L}$ & $\mathcal{G}$ & $\mathcal{R}$ & $\mathcal{L}$ & $\mathcal{G}$ & $\mathcal{R}$ \\
  \midrule
  \multirow{2}{*}{Indoor RF}
& Eval Grid
& \bnum{-2.7224} & \bnum{15.22} & \textbf{41.59\%}
& \num{-0.0594} & \num{1.06} & 0.91\%
& \num{-0.3030} & \num{1.35} & 4.63\%
& \num{-0.3778} & \num{1.46} & 5.77\% \\
& Eval Random
& \bnum{-3.3798} & \bnum{29.37} & \textbf{48.93\%}
& \num{-0.1190} & \num{1.13} & 1.72\%
& \num{-0.2650} & \num{1.30} & 3.84\%
& \num{-0.5078} & \num{1.66} & 7.35\% \\
  \bottomrule
\end{tabular}%
}
\end{table*}

\subsection{Maximum-Likelihood Observation Estimator}
\label{sec:estimator}
For channel estimation, we instantiate
the online estimator $\wh{\Psi}$ using an implementation of \eqref{eq:arg_max_sc} based on the iterative LM algorithm \cite{Bomfin_multiband}, which is known to be efficient at high SNR.
Let $\mathbf{b}_{\rm{F}}(\cdot)$ and $\mathbf{b}_{\rm{R}}(\cdot)$ be the frequency- and spatial-domain basis functions
\begin{align}
 &\mathbf{b}_{\rm{F}}(\tau) = \left(e^{-j 2\pi f_0 \tau \cdot m}\right)_{m=0}^{M_{\rm{F}}-1}, \\
 &\mathbf{b}_{\rm{R}}(\theta) = \left(e^{ -j \pi \sin \theta \cdot m }\right)_{m=0}^{M_{\rm{R}}-1},
\end{align}
where $M_{\rm{F}}$ is the number of subcarriers, $f_0$ is the subcarrier spacing, and $M_{\rm{R}}$ is the number of half-wavelength-spaced receive-ULA elements.
The frequency-space channel response over $K$ paths is
\begin{equation}
    {\bs s}({\Psi}) = \sum_{k=1}^{K} \mathbf{b}_{\rm{R}}(\theta_k) \otimes \mathbf{b}_{\rm{F}}(\tau_k)\alpha_k.
\end{equation}
Assuming equal-power pilots across subcarriers, the channel observation can be modeled as
\begin{equation}
   {\bs y_m} = {\bs s}({\Psi})  + {\bs w_m}, \label{eq:ym_model}
\end{equation}
where ${\bs w_m}$ is an additive white Gaussian noise (AWGN) vector whose entries have variance $\sigma^2$.
Given the likelihood under AWGN
\begin{equation}
    p({\bs y_m} \mid {\Psi}) \propto
    e^{
    -\frac{({\bs y_m} - {\bs s}({\Psi}))^{H}({\bs y_m} - {\bs s}({\Psi}))}{\sigma^2}
    },
\end{equation}
the maximum-likelihood estimate $\wh{\Psi}$ is defined as
\begin{equation}
   \wh{\Psi} = \operatorname*{arg\,max}\limits_{{\Psi}} p({\bs y_m} \mid {\Psi}).
    \label{eq:arg_max_sc}
\end{equation}

\subsection{Comparison Benchmarks}
We compare against three other benchmarks.
These baselines are visualized alongside LOCUS-DT in
\Cref{fig:dmc_fig2_methods}.
\textbf{Gauss-Cart} is a Gaussian posterior distribution with the transmitter location $\bs{x}^t$
in Cartesian coordinates.  In this baseline,
the score function is realized through a mean $\mu \in \R^d$ and covariance matrix $\bs{Q} \in \R^{d \times d}$, both of which are functions of the same full inputs as the LOCUS-DT estimator.
\textbf{Gauss-Polar} is the identical estimator,
but with the posterior in polar coordinates.
Finally, \textbf{GMM-Cart} uses a mixture with $k=3$ Gaussian components.

\subsection{Training and Evaluation Protocol}

All trainable models are optimized with AdamW for $2000$ epochs with weight
decay $10^{-4}$ and hidden-layer dimensions $(64,16)$.
LOCUS-DT uses the pairwise feature construction
whereas the Gaussian benchmarks use candidate-independent conditioning on the
receiver pose and IQ observation.
We use a learning rate of $5 \times 10^{-3}$ for LOCUS-DT and
$2 \times 10^{-3}$ for the Gaussian baselines.
All training runs are performed on a single NVIDIA A100 GPU.
LOCUS-DT requires approximately $2.5$ GPU-hours, Gauss-Cart $1.5$
GPU-hours, Gauss-Polar $1.5$ GPU-hours, and GMM-Cart $2.0$ GPU-hours, for a
total of approximately $7.5$ GPU-hours for the full benchmark suite.

Besides the training cross-entropy in \eqref{eq:mirror_ce_index}, we report the
adjusted loss
\begin{equation}
\mathcal{L}
=
\frac{1}{M}
\sum_{m=1}^{M}
\left(
-\log \hat p_{m c_m^\star}
- \log N_m
\right),
\label{eq:exp_adjusted_loss}
\end{equation}
which measures gain relative to uniform random guessing over the candidate set.
We further report the derived quantities
\begin{equation}
\mathcal{G} = e^{-\mathcal{L}},
\qquad
\mathcal{R} = -\frac{\mathcal{L}}{\log N_{\mathrm{cand}}},
\label{eq:exp_gain_ratio}
\end{equation}
where $N_{\mathrm{cand}}$ is the split-specific candidate count. Lower
$\mathcal{L}$ and higher $\mathcal{G}$ and $\mathcal{R}$ indicate better
performance; \Cref{tab:dmc-eval-baselines-vs-mirror} reports the comparison.

\subsection{Qualitative Comparison on Shared Tasks}

\Cref{fig:dmc_fig1_direct} shows LOCUS-DT on three distinct random layouts;
each row pairs a Sionna room with posterior heatmaps for six TX/RX
configurations. \Cref{fig:dmc_fig2_methods} compares LOCUS-DT with Gauss-Cart,
Gauss-Polar, and GMM-Cart on the same tasks and geometries. LOCUS-DT yields the
most localized and geometry-consistent posteriors, with high-probability
regions aligned with the true TX and wall- and reflection-induced structure.
Gauss-Cart and Gauss-Polar produce smooth single-mode fields that miss sharp
spatial discontinuities and multimodal ambiguity. GMM-Cart captures limited
multimodality, but its peaks remain broader and less aligned because it lacks
candidate-wise DT matching. \Cref{tab:dmc-eval-baselines-vs-mirror} confirms
that LOCUS-DT outperforms all Gaussian baselines on both evaluation sets.

%% file: conclusion.tex
\section{Conclusions} We presented LOCUS-DT for posterior transmitter
localization via DT-conditioned multipath matching. In simulation, it captures
reflection-induced multimodal ambiguity, outperforms Gaussian and
Gaussian-mixture baselines, and generalizes to unseen layouts. Future work will
evaluate greater DT mismatch and measured rooms reconstructed using SLAM.